\documentclass[%
 aip,
 amsmath,amssymb,
 reprint,%
]{revtex4-1}

\usepackage{graphicx}
\usepackage{dcolumn}
\usepackage{bm}

\usepackage[utf8]{inputenc}
\usepackage[T1]{fontenc}
\usepackage{mathptmx}
\usepackage{etoolbox}

\makeatletter
\def\@email#1#2{%
 \endgroup
 \patchcmd{\titleblock@produce}
  {\frontmatter@RRAPformat}
  {\frontmatter@RRAPformat{\produce@RRAP{*#1\href{mailto:#2}{#2}}}\frontmatter@RRAPformat}
  {}{}
}%
\makeatother
\begin{document}

\preprint{AIP/123-QED}

\title{Super-sensing: 100-Fold enhancement in THz time-domain spectroscopy contrast via superoscillating waveform shaping}
\author{Peisong Peng}
 
\author{Dusty R. Lindberg}%

\author{Gerard McCaul}%

\author{Denys I. Bondar}%

\author{Diyar Talbayev}%
 \email{dtalbaye@tulane.edu.}%

\affiliation{ 
Department of Physics and Engineering Physics, Tulane University, 6400 Freret St., New Orleans, 70118, USA 
}%

\date{\today}

\begin{abstract}
Superoscillations, where a band-limited wave oscillates locally faster than its highest Fourier harmonic, result from destructive interference between the harmonics. Here, we demonstrate that superoscillations enable a 100-fold contrast enhancement of terahertz wave passing through two similar samples. By optimizing the time-domain contrast within a short observation window through relative phase adjustments of the fundamental harmonics, we achieve maximum contrast when wave intensity is minimized locally, leading to superoscillations. This enhancement is observed in both simulations with Gaussian harmonics and experiments with quasi-sinusoidal terahertz harmonics, promising advanced terahertz imaging applications  in medicine, pharmaceuticals, stand-off hazard detection, and nondestructive evaluation. 
\end{abstract}

\maketitle

\section{\label{sec:level1}Introduction}

Beginning with Newton, who used prisms to split the colors of sunlight, the science of light has now become ubiquitous in the modern world. There is hardly a single field that does not employ optics in some capacity. Beyond its usual setting in physics, chemistry, and biology,  optical sensing has found a home in disciplines as diverse as art history \cite{Krgener2015} and archaeology \cite{Liggins2022}. A rather recent development in the science of light is the discovery of the phenomenon known as \textit{superoscillation}, where a band-limited signal oscillates faster than its fastest Fourier component within a particular spatial or temporal window \cite{berry2019roadmap, rogers_realising_2020}.  The experimental realization of the superoscillations in the spatial domain has led to the development of a new sub-wavelength optical microscopy method, where a bright superoscillatory hot spot in a dark field is created to achieve the super-resolution\cite{rogers2012super,rogers2013optical,rogers2018optimising,cheng2022super,li2018achromatic}.

By contrast, the experimental implementation of optical superoscillations in the time domain remains in its infancy, despite potential applications in optical pulse compression beyond the Fourier limit\cite{wong_temporal_2011} and their demonstrable improvement to range resolution\cite{wong_superoscillatory_2012, jordan_fundamental_2023, howell_super_2023}. Other promising phenomenology includes $super$-$transmission$ $-$ the delivery of high-frequency optical fields  beyond their propagation length in an absorbing medium\cite{eliezer2014super}. It was also proposed that superoscillations could manifest as resonant absorption outside the signal bandwidth\cite{kempf_driving_2017}. Despite this range of exploitable behaviour, experimental studies of superoscillations in the time domain have been rare. The phenomenon was studied at radiofrequencies\cite{wong_temporal_2011, wong_superoscillatory_2012, zarkovsky_transmission_2020, jordan_fundamental_2023, howell_super_2023}; at optical frequencies, superoscillating laser pulse envelopes have been constructed\cite{Eliezer:18} and were used to improve the temporal resolution in time-resolved optical experiments\cite{eliezer2017breaking}. Sub-terahertz superoscillations have been found in the time-evolution of optically-excited coherent acoustic phonons\cite{brehm_temporal_2020}.  Our group recently synthesized time-domain optical superoscillations at terahertz (THz) frequencies\cite{mccaul153803}. It is now possible to study experimentally the properties of these optical waveforms and their interaction with matter.

\begin{figure*}[pt]
    \centering
    \includegraphics[width=1.0\textwidth]{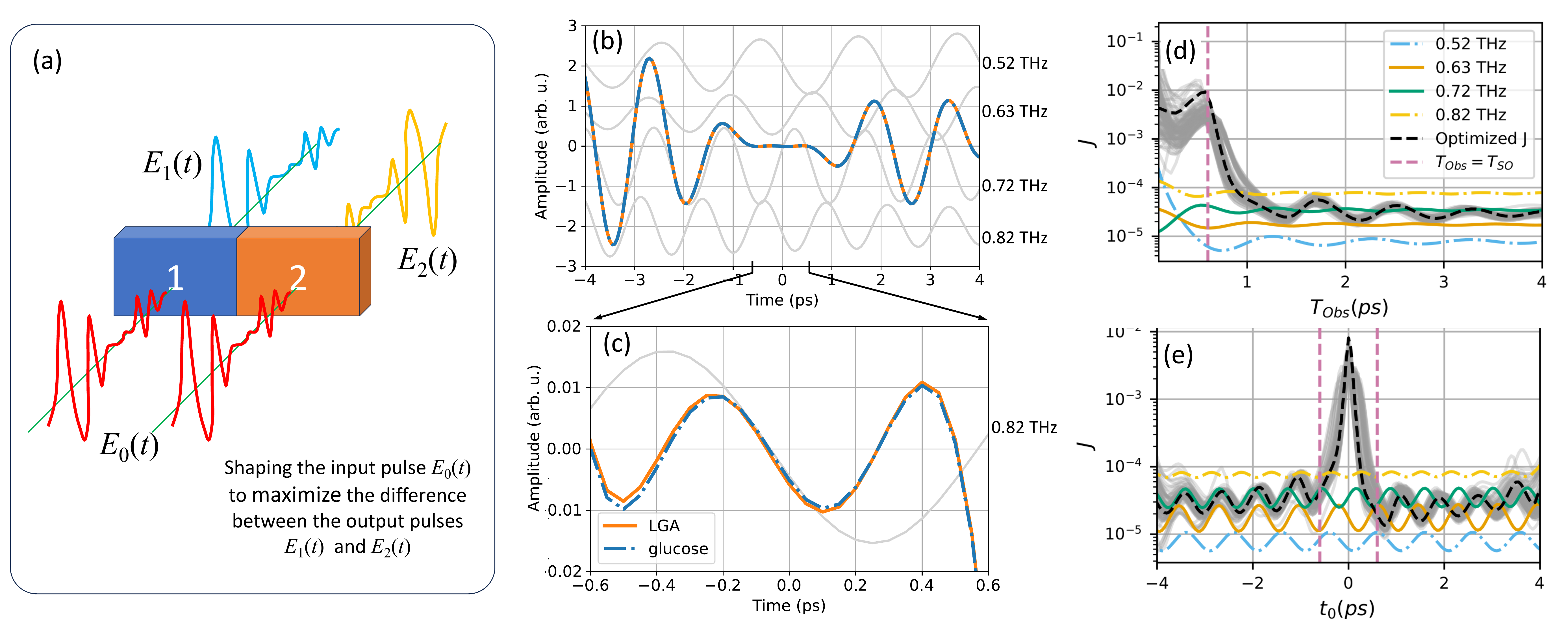}
    \caption{
    (a) A schematic of the experiment: we shape the input optical field $E_0(t)$ to maximize the contrast between the output fields $E_1(t)$ and $E_2(t)$ that pass through the optically similar samples 1 and 2.
    (b)-(e) Results of the numerical simulations:
    (b) Time-domain transmitted fields $E_1(t)$ and $E_2(t)$ that maximize the contrast $J$. The difference between $E_1(t)$ and $E_2(t)$ is not visible on this scale. The four quasi-sinusoidal harmonics that make up the full transmitted field are shown in gray together with their center frequencies.
    (c) A zoom-in of the $[-0.6, 0.6]$ ps time interval from panel (b) shows the small difference between $E_1(t)$ and $E_2(t)$. The 0.82-THz fundamental harmonic of the input field, which is scaled down to 2\% of its original amplitude is shown in gray. It is clear that the fields $E_1(t)$ and $E_2(t)$ are superoscillations, as they locally display shorter periods than the period of the 0.82-THz harmonic.
    (d) Contrast $J$ as a function of the observation window length $2\,T_{\rm obs}$. The observation window is always centered at the $t=0$ ps point in panels (b,c). The vertical dashed line marks $2\,T_{\rm obs}=1.2$ ps.
    (e) Contrast $J$ as a function of $t_0$, the center of the observation window of the fixed, 1.2-ps length. The vertical dashed lines indicate $t_0 = \pm 0.6$ ps. In both (d) and (e), the black dashed line represents the best result of optimization, and the gray lines are the 100 next best spectra predicted by our algorithm.
    }
    \label{fig:global1}
\end{figure*}

In this article, we examine the application of time-domain THz  superoscillations to linear optical sensing. THz optical sensing and imaging have been fruitfully employed in a number of fields\cite{el2013review, pickwell2006biomedical, yu2019medical, bessou2012advantage}. For example, it has been demonstrated that THz spectroscopy can be used to characterize and discriminate human cancer cell lines\cite{cao2021characterization}, while both transgenic soybean seeds \cite{liu2016discrimination} and rice \cite{xu2015discrimination} can be distinguished from their more conventional cousins using THz light. In biomedical applications, THz imaging of skin burns can be used to accurately predict wound healing outcomes \cite{khani_accurate_2022, khani_triage_2023}. THz imaging also allows the mapping of corneal hydration, which can be used to detect increased intraocular pressure \cite{chen_non-contact_2021}. In such sensing and imaging applications, we exploit the light wave's ability to provide contrast between two samples with similar optical properties. This contrast is a function of \textit{both} the samples' optical properties and the light used to distinguish them. We therefore ask the following question: in the scenario illustrated by Fig.~\ref{fig:global1}(a), how should the input optical waveform $E_0(t)$ be shaped such that the maximum contrast is achieved between each of two samples' output waveforms $E_1(t)$ and $E_2(t)$?  
In this optical setup, the input wave $E_0(t)$ is constructed as an interference waveform combining several fundamental quasi-sinusoidal harmonics with the possibility to set arbitrary relative amplitudes and phases between them\cite{mccaul153803}, and the wave transmission is governed solely by the linear optical response of the samples.

We find that the waveform that maximizes the contrast between optically similar samples is a superoscillation. We find a 100-fold increase in the optical contrast both in the numerically simulated and in the experimentally measured propagation of superoscillations through the two samples.  This superoscillation advantage is achieved in a short observation window in the time domain, shorter than some of the longer periods of the fundamental harmonics. Both these findings translate directly into significant advances in THz time-domain imaging technology.

\section{Results}
What is the waveform that produces the best contrast between two similar optical samples using a band-limited optical input field, Fig.~\ref{fig:global1}(a)? To answer this question, we need to quantify the contrast between similar samples. In our previous work\cite{mccaul153803}, we defined a contrast $J$ that characterizes the normalized difference between the time-domain spectra of two samples:
\begin{equation}
    J =  \frac{\int_{t_0-T_{\rm obs}}^{t_0+T_{\rm obs}}\left(E_1(t) - E_2(t)\right)^2dt}{\frac{1}{2}\int_{t_0-T_{\rm obs}}^{t_0+T_{\rm obs}}(E_1(t)^2 + E_2(t)^2)dt },
    \label{eq: normal contrast}
\end{equation}
where $E_{1,2}(t)$ denote the time-domain waveforms transmitted by the two samples (given the same input waveform $E_0(t)$). The interval $[t_0 - T_{\rm{obs}}, t_0 + T_{\rm{obs}}]$ represents the observation window in the time domain over which the two samples are compared; $t_0$ is the center point of the observation window and $2T_{\rm obs}$ is its length.

\subsection{Numerical Simulation}
We first explore this question through a numerical simulation. Two nonmagnetic (magnetic permeability $\mu = 1$) substances with similar dielectric permittivities will serve as samples. We model their dielectric permittivities $\varepsilon_1(\omega)$ and $\varepsilon_2(\omega)$ as
\begin{equation}
    \varepsilon_{1,2}(\omega) = \varepsilon_{\infty} + \frac{\omega_p^2}{(\omega_{1,2})^2-\omega^2-i\gamma\omega} 
    \label{eq: dielectric function},
\end{equation}
where the resonance frequencies $\omega_1 = 1.2$ THz and $\omega_2 = 1.4$ THz are the only parameters that distinguish the samples.  The remaining parameters are all the same for both samples: $\gamma = 0.1$ THz, $\omega_p = 0.05$ THz, and $\varepsilon_\infty = 2.3$. The sample thicknesses are set to 1 mm.  For the input fundamental harmonics, we choose $E_{0,j}(t) = \cos(\omega_j t)e^{-t^2/t_d^2}$ with frequencies $\{\omega_j\}_{j=1}^4 = \{0.52, 0.63, 0.72, 0.82\}$ THz and  $t_d = 20$ ps, which corresponds to the Gaussian envelope's full-width half-maximum of 33.3 ps. We choose these four fundamental harmonics to match the frequencies of our experimental setup. This simulation configuration ensures that the two samples have very similar optical properties with respect to the frequencies of the fundamental harmonics.

To get the transmitted fundamental harmonics, we first obtain the Fourier transform of $E_{0,j}(t)$, which we denote as $E_{0,j}(\omega)$. The transmitted harmonics are then determined by calculating the transmitted fields $E_{1,j}(\omega)$ and $E_{2,j}(\omega)$ in the frequency domain with the help of linear transmission coefficients $T_{1,2}(\omega)$\cite{duvillaret1996reliable}.  We then Fourier-transform the transmitted fields $E_{1,j}(\omega)$ and $E_{2,j}(\omega)$ back into the time domain to obtain the transmitted harmonics $E_{1,j}(t)$ and $E_{2,j}(t)$. The final interferometric waveforms are constructed as
\begin{eqnarray}
    \label{eq. out field 1}
    E_{1}(t)=\sum_{j=1}^{4}E_{1,j}(t-\tau_j),
    \label{eq. trans field 1}    
\end{eqnarray}
with a similar equation for the transmitted field $E_2(t)$. Here we introduce the possibility of varying the time delays $\tau_j$ between the fundamental harmonics, which must be the same for the incident and both transmitted fields. We shape the incident and transmitted fields in both the simulation and the experiment by varying the delays $\tau_j$.

With these preliminaries established, we numerically investigate the characteristics of the waveform that results from maximizing the contrast defined in Eq.~\eqref{eq: normal contrast}. This is achieved by varying the times delays (or relative phases) between the fundamental harmonics in Eq. \eqref{eq. out field 1}. (A detailed algorithm for finding the optimal time delays $\tau_j$ that maximize contrast $J$ can be found in the supplementary material.) The observation window is set with $t_0 = 0$ ps and $T_{\rm obs} = 0.6$ ps: the $[-0.6, 0.6]$ ps interval. The transmitted waveforms $E_1(t)$ and $E_2(t)$ that maximize $J$ in this window are shown in Fig.~\ref{fig:global1}(b).  This figure also shows the fundamental harmonics that make up the two transmitted waveforms. The difference between $E_1(t)$ and $E_2(t)$ is not visible on the scale of Fig.~\ref{fig:global1}(b).  To discern this difference, Fig.~\ref{fig:global1}(c) shows the zoomed-in view of the $[-0.6, 0.6]$ ps interval from Fig.~\ref{fig:global1}(b).

The most immediate finding of Fig.~\ref{fig:global1}(b) is that the transmitted intensity in the observation window is highly suppressed to near zero. Because the intensity of each individual fundamental harmonic remains unchanged during our numerical search for the maximum-$J$ waveform, the suppressed intensity is the result of the destructive interference between the harmonics.  It appears that the maximization of $J$ leads to the interferometric suppression of the intensity in the observation window. Figure~\ref{fig:global1}(c) compares the transmitted fields $E_1(t)$ and $E_2(t)$ to the 0.82 THz fundamental harmonic. The local oscillation period of $E_1(t)$ (and $E_2(t)$) in the observation window is approximately 0.6 ps, which is significantly shorter than the period of the 0.82 THz harmonic, which is equal to 1.2 ps. Therefore, both transmitted waveforms display \textit{superoscillations} within the $[-0.6, 0.6]$ ps time interval, Fig.~\ref{fig:global1}(c). The superoscillations, i.e., electric field oscillations faster than the fastest, 0.82-THz fundamental harmonic, exist only within the limited time window and only when the intensity is suppressed within that window due to the destructive interference between all fundamental harmonics. These conditions are the two hallmarks of superoscillations\cite{berry2019roadmap, rogers_realising_2020}. Thus, the waveforms $E_1(t)$ and $E_2(t)$ that maximize the contrast $J$ in the observation window exhibit superoscillatory behavior.

\begin{figure}[t]
    \centering
    \includegraphics[width=0.4\textwidth]{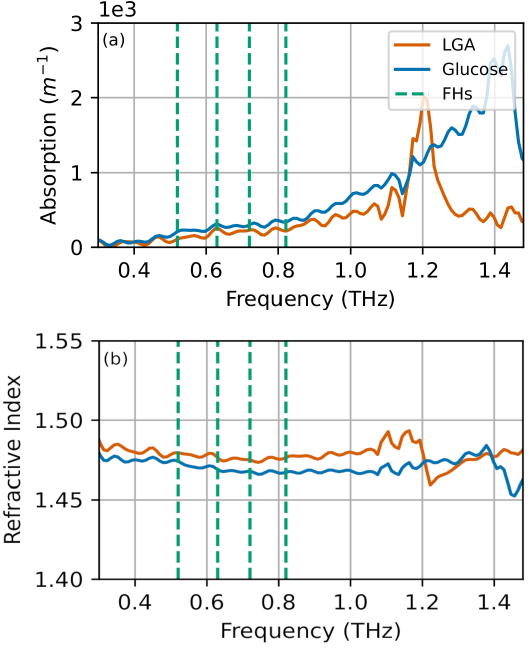}
    \caption{The (a) absorption and (b) refractive index measured using the broadband THz time-domain spectrometer for \emph{$\alpha$-d-Glucose} and \emph{L-Glutamic Acid} (LGA). The frequencies of the fundamental harmonics (FHs) used for contrast optimization, 0.52 - 0.82 THz, are shown as vertical dashed lines. In the region of the harmonics, the glucose and LGA samples have very similar absorption coefficient and refractive index.}
    \label{fig: index and absorption}
\end{figure}

We now compute the contrast for the superoscillating waveforms $E_1(t)$ and $E_2(t)$ from Figs.~\ref{fig:global1}(b,c) as a function of the observation window length $T_{\rm{obs}}$ and show it in Fig.~\ref{fig:global1}(d). For long observation windows, the contrast is comparable to that of the fundamental harmonics.  When the observation window becomes shorter than $2$ ps, the contrast grows rapidly and reaches a maximum for observation windows equal to or shorter that $2\,T_{\rm obs} = 1.2$ ps, which corresponds to the interval with the superoscillating behavior. Fig.~\ref{fig:global1}(d) illustrates the superoscillation contrast advantage exceeding two orders of magnitude for these observation windows.

Another way to compare the contrast $J$ of the superoscillating waveforms to the fundamental harmonics is to `translate' the observation window along the transmitted waveforms $E_1(t)$ and $E_2(t)$ over the whole range of the horizontal time axis in Fig.~\ref{fig:global1}(b). We perform this `translation' by varying the center point $t_0$ of the observation window $[t_0 - T_{\rm{obs}}, t_0 + T_{\rm{obs}}]$ while keeping the fixed $T_{\rm{obs}} = 0.6$ ps.  We compute $J$ as the center point $t_0$ varies from -4 ps and 4 ps. The resulting contrast $J$ is plotted as the black dashed line in Fig.~\ref{fig:global1}(e), where the horizontal axis directly corresponds to the horizontal axis of panel (b).  We see that $J$ sharply peaks within the superoscillating region of $E_1(t)$ and $E_2(t)$ ($[-0.6, 0.6]$ ps) and is mostly comparable to $J$ of the fundamental harmonics everywhere else. In this instance the contrast advantage of the superoscillating waveform over the fundamental harmonics also approaches two orders of magnitude.

\begin{figure*}
    \centering
    \includegraphics[width=0.8\textwidth]{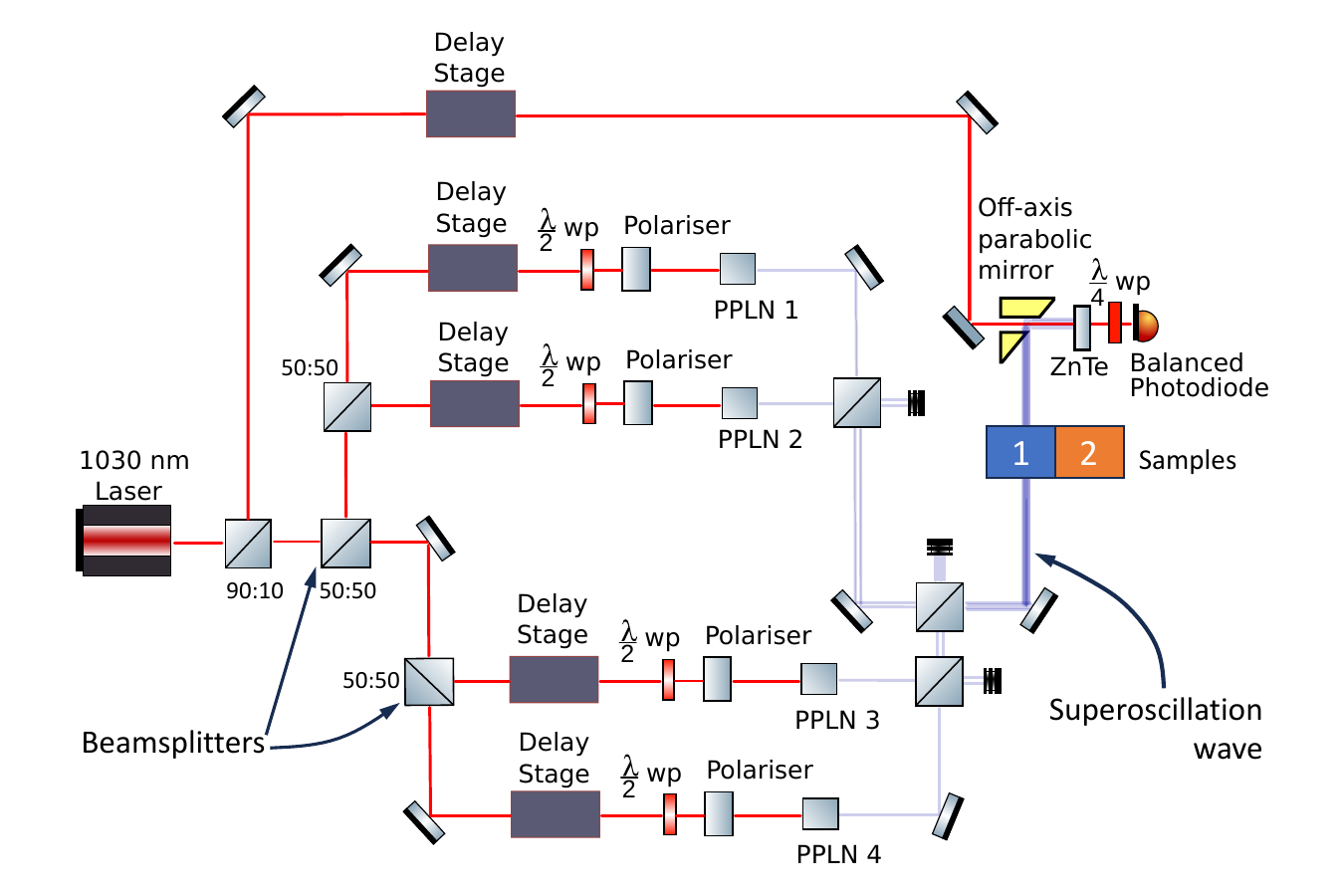}
    \caption{Schematic layout of the superoscillation synthesis setup. Wp - waveplate. PPLN - periodically poled lithium niobate crystal. }
    \label{fig:setup}
\end{figure*}

\subsection{Experimental Demonstration}
We have experimentally implemented the described protocol for maximizing the contrast between two similar samples in a time-domain THz spectroscopic measurement.  We chose polytetrafluoroethylene (PTFE) pellets with small amounts of mixed-in molecular crystal powders as samples.  Pure PTFE pellets possess a refractive index of approximately 1.5 and no resonance features below 2 THz at room temperature\cite{chantry1974far}.  We used the L-Glutamic acid (LGA) and the anhydrous $\alpha$-D-Glucose (glucose) as the molecular materials that make our two pellets display slightly different THz optical properties.  At room temperature, LGA crystals display a vibrational THz absorption resonance near 1.2 THz, while the lowest vibrational resonance of glucose is found near 1.4 THz\cite{taday2003terahertz, da2022analysis}.  These materials also motivated the selection of resonance frequencies $\omega_1 = 1.2$ THz and $\omega_2 = 1.4$ THz in our computational model in Eq.~\eqref{eq: dielectric function}. We prepared PTFE pellets containing 5\% by mass of either glucose or LGA.  The pellets were compressed under the pressure of $10$ Mpa. The resulting pellets showed diameters of 10 mm and thicknesses of 1.1 mm.  We confirmed that the two pellets exhibit very similar optical properties using broadband THz time-domain spectroscopy, Fig.~\ref{fig: index and absorption}.

Fig.~\ref{fig:setup} shows the schematic of the experimental setup for the synthesis of the superoscillations and the study of the superoscillation contrast between samples 1 and 2. An infrared femtosecond laser beam (167 fs, 1030 nm center wavelength) is split into four THz generation beams and one optical gating beam. Each of the split beams passes through an optical delay stage that controls the relative phases (time delays) between them. Half-waveplate-and-polarizer pairs in the THz generation beams allow the control of the THz generation beam intensity. The narrow-band, quasi-sinusoidal THz waves are emitted in the 5-mm long and 0.5 mm thick periodically poled lithium niobate crystals (PPLNs) with appropriate poling periods to generate the 0.52 THz, 0.63 THz, 0.72 THz, and 0.82 THz harmonics\cite{lee_generation_2000, lee_temperature_2000, weiss2001tuning, lhuillier_generation_2007, lhuillier_generation_2007-1}. (Detailed time- and frequency-domain spectra of each harmonic can be found in the supplementary material.) The emitted THz harmonics are combined in a single THz superoscillation beam by beamsplitters and pass either through sample 1 or sample 2. The relative phases between the fundamental THz harmonics are controlled by optical delay stages in the path of the femtosecond optical pulses \cite{mccaul153803}. The transmitted time-domain electric field is detected by electro-optic sampling in a ZnTe crystal by the optical gating beam\cite{wu_freespace_1995, nahata_wideband_1996, lin_measurement_2018}.

\begin{figure}
    \centering
    \includegraphics[width=0.5\textwidth]{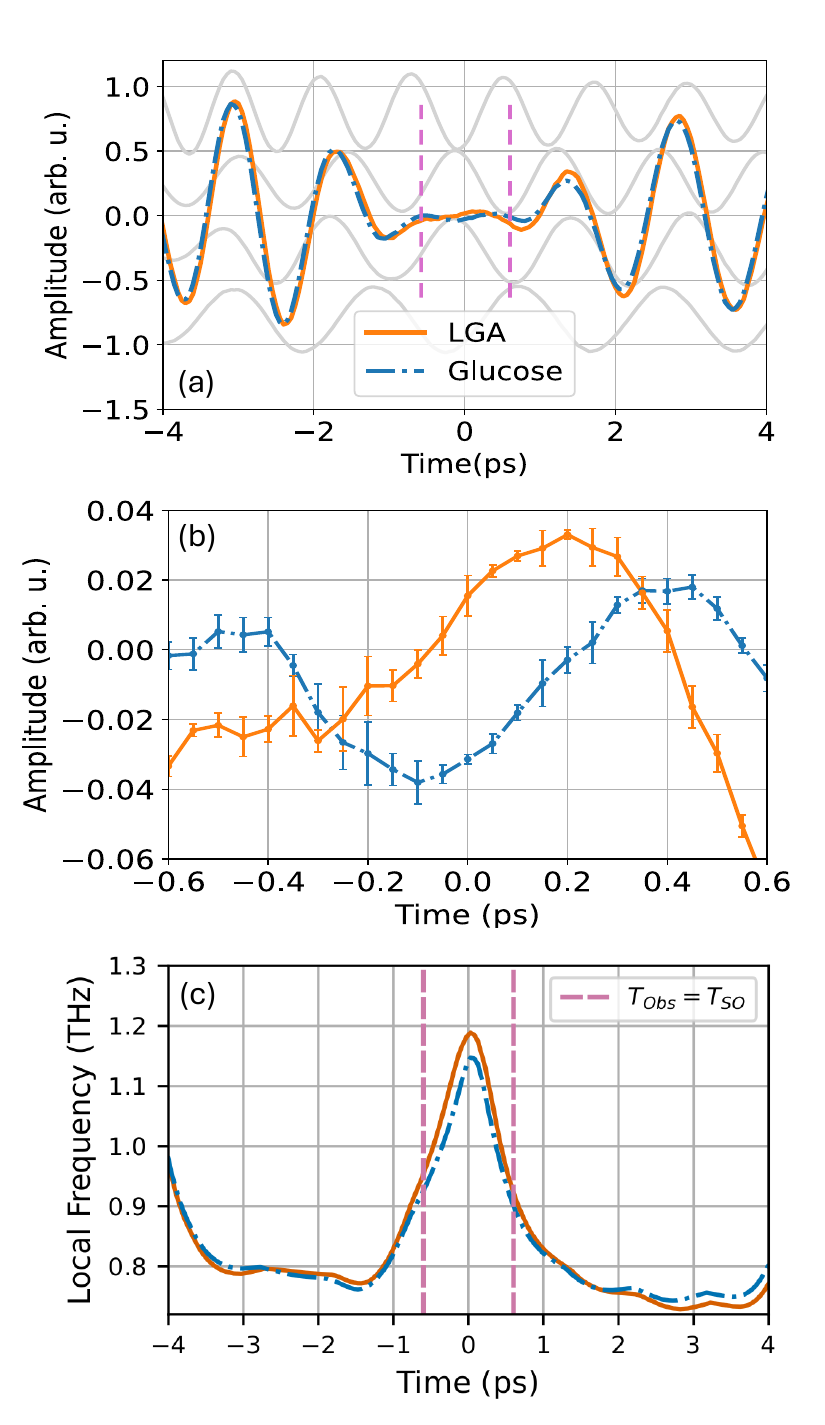}
    \caption{(a) Measured transmitted waveforms $E_{\rm{g}}(t)$ and $E_{\rm{L}}(t)$ for the glucose and LGA samples. The vertical dashed lines indicate the observation window where the contrast $J$ is maximized. The fundamental harmonics that constitute the glucose waveform $E_{\rm{g}}(t)$ are shown in the background in grey: 0.52 THz, 0.63 THz, 0.72 THz, and 0.82 THz from bottom to top. (b) Zoomed-in view of the $[-0.6,0.6]$ ps intevral from panel (a). (c) The local frequency calculated using the short-time Fourier transform method for $E_{\rm{g}}(t)$ and $E_{\rm{L}}(t)$. The vertical dashed lines indicate the observation window.}
    \label{fig: experimental}
\end{figure}

To implement the maximum contrast protocol, we start with measuring the transmitted individual fundamental harmonics for the glucose and LGA samples. We use these measured harmonics as inputs to the same algorithm that maximizes $J$ in the simulation described above. We first numerically add the harmonics to form transmitted interference waveforms $E_{\rm{g}}(t)$ and $E_{\rm{L}}(t)$ for glucose and LGA, respectively. We use these two waveforms to compute the contrast $J$ according to Eq.~\eqref{eq: normal contrast} and optimize it by varying the relative time delays between the measured harmonics as they combine to form the fields $E_{\rm{g}}(t)$ and $E_{\rm{L}}(t)$.  The observation window parameters are $t_0=0$ ps and $T_{\rm{obs}} = 0.6$ ps; the window length is $2T_{\rm{obs}} = 1.2$ ps. This procedure produces a set of time delays $\{\tau_j^{\rm{exp}}\}_{j=1}^4$ that maximize $J$. We then add the resulting time delays $\{\tau_j^{\rm{exp}}\}_{j=1}^4$ to each incident fundamental harmonic in the experiment, combine them into one THz beam, and record the transmitted interference waveforms $E_{\rm{g}}(t)$ and $E_{\rm{L}}(t)$ for the glucose and LGA samples. The recorded experimental waveforms are shown in Fig.~\ref{fig: experimental}(a), where the observation window is indicated by the vertical dashed lines. A more detailed data set representing each harmonic and full waveforms for both samples before and after applying the optimal time delays $\{\tau_j^{\rm{exp}}\}_{j=1}^4$ can be found in the supplementary material.

We find that maximization of the contrast $J$ in the experiment greatly suppresses the interferometric intensity in the observation window, Fig.~\ref{fig: experimental}(a).  This is the same outcome that we found in the numerical simulation section. Unlike the simulation, the experimental data contain noise, and the suppression of interferometric intensity may degrade the measurement of the contrast $J$ between the samples. Figure~\ref{fig: experimental}(b) shows a zoomed-in view of the measured transmitted fields $E_{\rm{g}}(t)$ and $E_{\rm{L}}(t)$ within the the $[-0.6,0.6]$ ps observation window together with the measurement error bars. We find that the signal level is comfortably above the noise level, thus confirming the reliability of the measurement of $J$ and of our analysis. We also point out that according to our definition of the contrast, Eq.~\eqref{eq: normal contrast}, simply increasing the overall intensity of the harmonics does not change $J$, as is it normalized by the sum of harmonic intensities.

Similar to our numerical simulations, we find that the experimental transmitted waveforms $E_{\rm{g}}(t)$ and $E_{\rm{L}}(t)$ also display superoscillations. Figure~\ref{fig: experimental}(b) shows the local frequency of the two waveforms calculated using the short-time-Fourier-transform method. The local frequency is higher for $E_{\rm{g}}(t)$ and $E_{\rm{L}}(t)$ than their corresponding fundamental harmonics with the 0.82 THz center frequency. This result confirms the superoscillatory nature of the transmitted waveforms in the observation window, Fig.~\ref{fig: experimental}(a); it also agrees very well with what we have seen in our numerical simulation above.

\begin{figure}
    \centering
    \includegraphics[width = 0.5\textwidth]{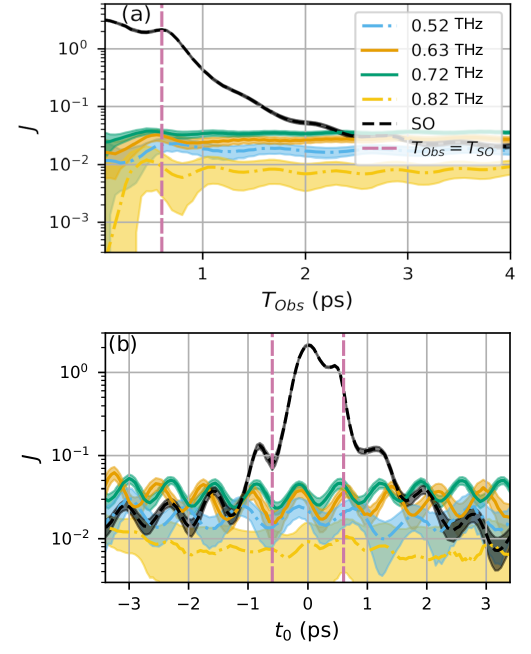}
    \caption{(a) Contrast $J$ as a function of the observation window length $T_{obs}$. The vertical dashed line marks where $T_{obs} = 0.6$ (ps), which is the length of the window of superoscillation, $T_{SO}$. The partially-transparent color bands around each graph indicate the experimental error bars. (b) Contrast $J$ as a function of $t_0$, the center of the observation window of the fixed, 1.2-ps length. The vertical dashed lines indicate $t_0 = \pm 0.6$ ps.
    }
    \label{fig: J experiment}
\end{figure}

In Fig.~\ref{fig: J experiment}, we compare the experimentally achieved contrast $J$ for the fundamental harmonics and the superoscillating fields $E_{\rm{g}}(t)$ and $E_{\rm{L}}(t)$.  Panel (a) shows the dependence of $J$ on the length of the observation window ($2\,T_{\rm{obs}}$); the center of the window remains fixed at the 0 ps time point of Fig.~\ref{fig: experimental}(a). The contrast advantage of the superoscillating waveforms over the fundamental harmonics approaches two orders of magnitude for observation windows shorter than 1.2 ps.  The advantage drops for longer windows and becomes comparable to the contrast of individual harmonics when $2\,T_{obs} > 4$ ps.  Figure~\ref{fig: J experiment}(b) shows the dependence of $J$ on the position of the center $t_0$ of the observation window along the transmitted waveforms; the length of the observation window is fixed at $2T_{\rm{obs}} = 1.2$ ps. We find a strong, almost 100-fold, enhancement of $J$ when the observation window is centered on the minimum of interferometric intensity at $t_0 = 0$ ps. When the observation window is centered on the interferometric maxima ($t_0 =$ -3 ps and $t_0 =$ 3 ps, Figs.~\ref{fig: experimental}(a) and~\ref{fig: J experiment}(b)), the transmitted waveforms $E_{\rm{g}}(t)$ and $E_{\rm{L}}(t)$ lose the contrast advantage over the individual fundamental harmonics. 

\section{Discussion}
We described the maximization of the time-domain contrast $J$ between two similar samples numerically using the Gaussian-pulse fundamental harmonics and experimentally using the quasi-sinusoidal harmonics emitted from PPLNs. In the simulation and in the experiment, we achieve 100-fold contrast enhancement when the fundamental harmonics add destructively and the transmitted intensity is minimized within the observation window $2\,T_{\rm obs}$ - the same window where the contrast $J$ is maximized. Returning to the question we asked at the beginning: is superoscillation the optimal waveform for maximizing the contrast between two optical samples? Our previous study of THz time-domain superoscillations showed that the minimization of interferometric intensity always results in superoscillatory fields \cite{mccaul153803}. Conversely, construction of a superoscillation results in a local intensity minimum due to the destructive interference between the Fourier components of a band-limited signal.  Since maximizing the contrast $J$ is equivalent to minimizing the transmitted interferometric intensity, we can reason that the time-domain superoscillation in the transmitted field is indeed the optimal waveform to achieve the maximum contrast.

\begin{figure}
    \centering
    \includegraphics[width = 0.5\textwidth]{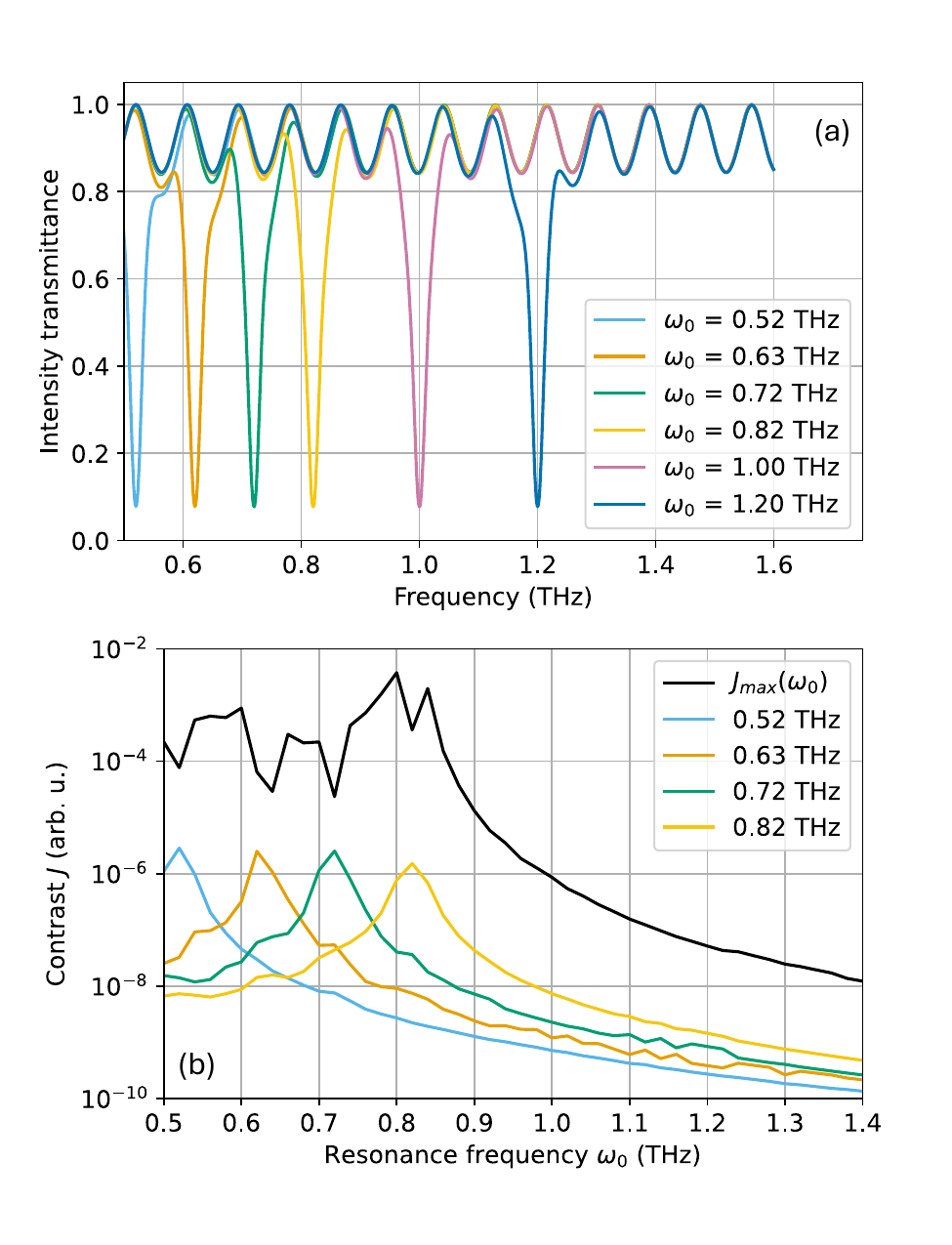}
    \caption{(a) Calculated intensity transmittance of the modeled samples for several representative resonance frequencies $\omega_0$. (b) Black line: Optimized contrast $J_{max}(\omega_0)$ of the superoscillating waveform as a function of resonance frequency $\omega_0$. Color lines: contrast achieved by individual fundamental harmonics $\{0.52, 0.63, 0.72, 0.82\}$ THz.
    }
    \label{fig:resonant_j}
\end{figure}

The described simulations and measurements featured optical resonances (1.2 THz and 1.4 THz) outside of the frequency band of the fundamental harmonics (0.5 - 0.8 THz). Such off-resonant sensing can be useful when samples or media are completely opaque due to absorption or when the resonant light wave can be harmful, e.g., ultraviolet light for biological specimens. Our results show that high-contrast sensing in these circumstances can be performed with off-resonant time-domain waveforms.  A question arises whether the described sensing procedure retains its advantages when the optical resonances of the medium fall within the frequency band of the fundamental harmonics. To answer this, we model the contrast $J$ between two samples with the same resonance frequency $\omega_0$. We use the dielectric function of Eq.~\eqref{eq: dielectric function} for the two samples. The oscillator strengths ($\omega_p^2$) are fixed but different for each sample -- they differ by 0.1\% -- which introduces a slight difference in their optical properties ($\omega_p = 0.05$ and $0.05 \times 1.001$ THz). The scattering rate and the high-frequency dielectric constant are set to $\gamma = 0.1$ THz and $\varepsilon_\infty = 2.3$. Fig.~\ref{fig:resonant_j}(a) illustrates the intensity transmittance of the samples as a function of $\omega_0$; the samples are transparent with the exception of the absorption at the resonance frequency, where they transmit only about 10\% of light. The two samples have almost identical properties, so only the transmittance spectrum of one sample is shown in panel (a). The transmitted optical waveforms are constructed using Eq.~\eqref{eq. out field 1} from the same four fundamental harmonics as before, $\{0.52, 0.63, 0.72, 0.82\}$ THz, and the contrast $J$ is maximized in the observation window $[-0.6, 0.6]$ ps by varying the phase delays $\tau_j$. We repeat this numerical maximization of $J$ for resonance frequencies $\omega_0$ in the $0.5 - 1.5$ THz range, which includes frequencies within and outside the frequency band of the four fundamental harmonics.

The black line in Fig.~\ref{fig:resonant_j}(b) shows the outcome of this numerical maximization of contrast $J$ as a function of $\omega_0$ -- $J_{max}(\omega_0)$. For $\omega_0 < 0.85$ THz, contrast $J_{max}(\omega_0)$ remains high but drops dramatically at higher frequencies. The contrast is clearly much higher when the resonance frequency $\omega_0$ falls within the frequency band of the sensing waveform. However, for all values of $\omega_0$ in the $0.5 - 1.5$ THz range, the optimal $J_{max}(\omega_0)$ is achieved when the transmitted waveforms display the minimum interferometric intensity in the $[-0.6, 0.6]$ ps observation window, thus leading to the emergence of superoscillations. We also compare the contrast $J_{max}(\omega_0)$ with the contrast $J$ achieved by using only the individual fundamental harmonics shown by colored lines in Fig.~\ref{fig:resonant_j}(b). When $\omega_0$ is outside the frequency band of fundamental harmonics ($\omega_0 > 0.85$ THz), $J_{max}(\omega_0)$ exceeds the contrast $J$ of the individual harmonics by two orders of magnitude, which is the same contrast advantage found in both the experiment and the simulations of the previous section. When $\omega_0$ matches one of the fundamental frequencies ($\{0.52, 0.63, 0.72, 0.82\}$ THz), the contrast advantage of $J_{max}(\omega_0)$ is diminished, but nevertheless exceeds one order of magnitude in each case. This result leads to an important conclusion: even when the incident wave of limited bandwidth is fully resonant with an optical transition, sensing contrast can be dramatically improved by superoscillating waveform shaping.

Another important question is whether the specific method used in this work to synthesize time-domain waveforms is essential for achieving the observed contrast enhancement from superoscillatory shaping. Our THz harmonics span a spectral range of 0.3 THz, yet time-domain THz sources with bandwidths ranging from 1 THz to tens of THz are now available \cite{leitenstorfer_2023_2023}. As well, several methods for shaping THz pulses have been demonstrated, including metasurface-based approaches \cite{keren-zur_direct_2019, veli_terahertz_2021}, manipulation of the generating infrared pulse \cite{liu_enhancement_1996, liu_terahertz_1996, sohn_tunable_2002, vidal_femtosecond_2010, ropagnol_thz_2011, sato_terahertz_2013}, and other techniques \cite{gingras_direct_2017}. Our central result -- the enhancement of time-domain contrast when the interferometric intensity of a band-limited waveform is suppressed -- remains valid regardless of the signal bandwidth or the specific pulse shaping method employed. We selected our approach because it offers exceptionally precise control over the relative phases and amplitudes of the constituent harmonics. For instance, relative phase shifts of tens of multiples of $2\pi$ can be accurately implemented. However, our method is neither compact nor easily scalable. Extending the bandwidth requires incorporating additional fundamental harmonics, which quickly becomes impractical beyond the four harmonics used here. Nonetheless, we believe the generality of our core finding implies that a wide range of THz generation and time-domain pulse shaping techniques can be used to harness the demonstrated contrast enhancement.

\section{Conclusions}
In this article, we investigate how to shape a band-limited optical pulse to maximize its ability to distinguish between two similar samples based on their linear optical responses. The incident wave consists of a finite number of fundamental harmonics and is shaped by varying the relative time delays between them. We find that the contrast $J$ is maximized in a short observation window $2\,T_{\rm obs}$ whose length is roughly equal to the oscillation period of the highest-frequency fundamental harmonic in the input pulse.  We also find that within this short observation window, the contrast is maximized when the transmitted waveform possesses a minimum of interferometric intensity - that is, the transmitted fundamental harmonics interfere destructively. Outside the observation window, the harmonics interfere constructively and the contrast advantage over the individual harmonics is lost.  We observe that the transmitted waveforms that maximize the contrast also display time-domain superoscillations.  This is consistent with our previous results, where we report that establishing destructive interference between harmonics produces superoscillations \cite{mccaul153803}.  In both experiment and numerical modeling, we found 100-fold enhancement in the contrast $J$ due to superoscillations compared to the individual fundamental harmonics.

Our findings suggest a number of imaging applications employing THz technology, where $J$ would serve as a measure of the image contrast. Such 100-fold contrast enhancement could vastly improve THz imaging, especially in the medical field where THz light is used for tumor diagnostics and visualization.  Potentially, the much higher contrast could contribute to earlier tumor detection.  We find that the contrast is maximized in a short time window, shorter than the oscillation period of most of the fundamental harmonics, meaning  \textit{the less we look, the more we see}. Furthermore, the image collection time in a time-domain measurement is directly proportional to the length of the observation window. By only recording the transmitted waveforms in the shortened observation window, we can drastically shorten the total time needed to construct the image. Limiting the total imaging time is important for in-vivo imaging, for example to minimize occlusion effects. These examples demonstrate that our fundamental results on the propagation of time-domain superoscillations can also lead to technological breakthroughs in THz time-domain imaging applications. 

\section*{Supplementary Material}
The supplementary material includes the following: time-domain THz waveforms emitted from each of the PPLNs and their frequency-domain amplitudes, a more detailed algorithm for maximizing the contrast $J$ between two theoretical or measured waveforms, and an example of applying the algorithm to the experimental harmonics to illustrate the resulting contrast enhancement.

\begin{acknowledgments}
The authors acknowledge the very generous support of the W. M. Keck Foundation. G.M. is supported by the European Research Council (ERC) under the European Union’s Horizon 2020 Research and Innovation Program (grant agreement 833365) D.I.B. is also supported by Army Research Office (ARO) (grant W911NF-23-1-0288; program manager Dr.~James Joseph). The views and conclusions contained in this document are those of the authors and should not be interpreted as representing the official policies, either expressed or implied, of ARO or the U.S. Government. The U.S. Government is authorized to reproduce and distribute reprints for Government purposes notwithstanding any copyright notation herein. 
\end{acknowledgments}

\section*{Data Availability Statement}

The data that support the findings of this study are available from the corresponding author upon reasonable request.

\bibliography{reference}

\end{document}